\documentclass[aps,prl,reprint,balancelastpage,nofootinbib,preprintnumbers,superscriptaddress]{revtex4-1}

\usepackage{slashed}
\usepackage{bm}
\usepackage{latexsym,amssymb,amsmath,float,url,mathrsfs}
\usepackage{latexsym}
\usepackage{graphicx}
\usepackage{epstopdf}
\usepackage{amsmath}
\usepackage{comment}
\usepackage{subfigure}
\usepackage{natbib}
\usepackage{hyperref}
\usepackage{pifont}
\usepackage{blindtext}
\usepackage{adjustbox}
\usepackage{multirow}
\usepackage{tabularx}
\usepackage[table]{xcolor}
\usepackage{orcidlink}
\usepackage{tikz}
\usetikzlibrary{tikzmark}
\usepackage{fontawesome}

\usepackage{blkarray}
\usepackage{graphicx}
\usepackage{amsfonts}
\usepackage{soul}
\usepackage{amssymb}
\usepackage{amsmath}
\usepackage{cancel}
\usepackage{tcolorbox}

\usepackage{graphics,appendix,afterpage,makecell} 

\definecolor{oucrimsonred}{rgb}{0.6, 0.0, 0.0}
\definecolor{persianblue}{rgb}{0.11, 0.22, 0.73}
\definecolor{forestgreen}{rgb}{0.13,0.35,0.13}
\definecolor{lightgray}{rgb}{0.83, 0.83, 0.83}
 \hypersetup{colorlinks, citecolor=oucrimsonred, linkcolor=black, urlcolor=oucrimsonred}
\definecolor{cornellred}{rgb}{0.7, 0.11, 0.11}
\definecolor{navyblue}{rgb}{0.0, 0.0, 0.5}
\definecolor{amethyst}{rgb}{0.6, 0.4, 0.8}
\definecolor{yellow}{rgb}{1.0, 1.0, 0.0}
\definecolor{firebrick}{rgb}{0.7, 0.13, 0.13}
\definecolor{tangerineyellow}{rgb}{1.0, 0.8, 0.0}
\definecolor{deepfuchsia}{rgb}{0.76, 0.33, 0.76}
\definecolor{amber}{rgb}{1.0, 0.75, 0.0}
\definecolor{VioletRed4}{rgb}{0.55, 0.13, .32}
\definecolor{indiagreen}{rgb}{0.07, 0.53, 0.03}
\definecolor{VioletRed4}{rgb}{0.55, 0.13, .32}
\newcommand{\be}{\begin{equation}}
\newcommand{\ee}{\end{equation}}
\newcommand{\bea}{\begin{equation} \begin{aligned}}
\newcommand{\eea}{\end{aligned} \end{equation}}

\definecolor{oucrimsonred}{rgb}{0.6, 0.0, 0.0}
\newcommand\vertarrowbox[3][6ex]{%
  \begin{array}[t]{@{}c@{}} #2 \\
  \left\uparrow\vcenter{\hrule height #1}\right.\kern-\nulldelimiterspace\\
  \makebox[0pt]{\scriptsize#3}
  \end{array}%
}

\definecolor{verdechiaro}{rgb}{0.6,1,0.6}
\definecolor{giallochiaro}{rgb}{1,1,0.6}
\definecolor{bluscuro}{rgb}{0.15, 0.2, 0.9}
\definecolor{verdes}{rgb}{0.1, 0.5, 0.1}%
\definecolor{tangerineyellow}{rgb}{1.0, 0.8, 0.0}

\definecolor{americanrose}{rgb}{1.0, 0.01, 0.24}
\definecolor{cobalt}{rgb}{0.0, 0.28, 0.67}
\definecolor{brandeisblue}{rgb}{0.0, 0.44, 1.0}
\definecolor{mycolor}{rgb}{0.0, 0.0, 0.5}
\definecolor{oxfordblue}{rgb}{0.0, 0.13, 0.28}
\definecolor{azure}{rgb}{0.0, 0.5, 1.0}
\definecolor{turquoiseblue}{rgb}{0.0, 1.0, 0.94}
\newtcolorbox{mynewbox}[1]{colback=white!5!white,colframe=azure!75!black,fonttitle=\bfseries,title=#1}
\newtcolorbox{mybox}{colback=mycolor!5!white,colframe=azure!75!black}
\newtcolorbox{mynamedbox}[1]{colback=mycolor!5!white,colframe=azure!75!black,title=#1}
\definecolor{venetianred}{rgb}{0.78, 0.03, 0.08}
\newtcolorbox{mynamedbox1}[1]{colback=venetianred!5!white,colframe=venetianred!80!black,title=#1}
\newtcolorbox{mynamedbox2}[1]{colback=azure!5!white,colframe=azure!80!black,title=#1}

\definecolor{verdes}{rgb}{0.1, 0.5, 0.1}%
\definecolor{cornellred}{rgb}{0.7, 0.11, 0.11}

\definecolor{VioletRed4}{rgb}{0.55, 0.13, .32}

\hypersetup{
     colorlinks   = true,
     citecolor    = violet,
     urlcolor     = violet,
     linkcolor    = violet}

\definecolor{rossocorsa}{rgb}{0.83, 0.0, 0.0}

\usepackage[normalem]{ulem}

\newcommand{\papertitle}{  Nonlinearities of   Schwarzschild Black Hole Head-on  Collisions}

\begin{document}

\title[]{\papertitle}

\author{Alex Kehagias\orcidlink{}}
\affiliation{Physics Division, National Technical University of Athens, Athens, 15780, Greece}
\affiliation{CERN, Theoretical Physics Department, Geneva, Switzerland}

\author{Davide Perrone\orcidlink{0000-0003-4430-4914}}
\affiliation{Department of Theoretical Physics and Gravitational Wave Science Center,  \\
24 quai E. Ansermet, CH-1211 Geneva 4, Switzerland}

\author{Antonio Riotto\orcidlink{0000-0001-6948-0856}}
\affiliation{Department of Theoretical Physics and Gravitational Wave Science Center,  \\
24 quai E. Ansermet, CH-1211 Geneva 4, Switzerland}


\begin{abstract}
\noindent
We derive analytically the amplitude of the quadratic quasi-normal mode generated in the ringdown stage of the gravitational waveform produced by the ultra-relativistic head-on collision of two non-spinning Schwarzschild black holes. 
Although being a highly nonlinear event,  second-order perturbation theory suffices and that nonlinearities may be derived by a simple bootstrapping procedure.

\end{abstract}

\maketitle

\noindent\textbf{Introduction --} 
The advent of gravitational wave (GW) astronomy has ushered in a new era for high-precision tests of strong-field gravity \cite{Berti:2015itd,Berti:2018vdi,Franciolini:2018uyq,LIGOScientific:2020tif}. 
Current ground-based interferometers—LIGO, VIRGO, and KAGRA—together with the forthcoming space-based mission LISA, are achieving sensitivities that enable detailed studies of the ringdown phase of black hole (BH) mergers \cite{Berti:2005ys,KAGRA:2013rdx,LIGOScientific:2016aoc,KAGRA:2021vkt,LIGOScientific:2023lpe}. 
A central objective of these efforts is to characterize BH perturbations at late times, where both linear and nonlinear effects leave distinct signatures in the emitted GWs.

Traditionally, our understanding of BH ringdowns has relied heavily on linear perturbation theory. 
Within this framework, BHs are expected to display exponentially damped oscillations—quasinormal modes (QNMs)—at intermediate times, followed by a late-time power-law decay \cite{Berti:2025hly}. 

Significant progress has been made towards refining predictions for both QNM spectra and late-time tails \cite{Ching:1994bd,Krivan:1996da,Krivan:1997hc,Krivan:1999wh,Burko:2002bt,Burko:2007ju,Hod:2009my,Burko:2010zj,Racz:2011qu,Zenginoglu:2012us,Burko:2013bra,Baibhav:2023clw,Rosato:2025rtr,Ianniccari:2025avm}. 
These linear predictions have been robustly validated through numerical relativity simulations and increasingly precise GW observations.

In recent years, growing attention has shifted toward nonlinear aspects of BH perturbations \cite{Gleiser:1995gx,Gleiser:1998rw,Campanelli:1998jv,Garat:1999vr,Zlochower:2003yh,Brizuela:2006ne,Brizuela:2007zza,Nakano:2007cj,Brizuela:2009qd,Ripley:2020xby,Loutrel:2020wbw,Pazos:2010xf,Sberna:2021eui,Redondo-Yuste:2023seq,Mitman:2022qdl,Cheung:2022rbm,Ioka:2007ak,Kehagias:2023ctr,Khera:2023oyf,Bucciotti:2023ets,Spiers:2023cip,Ma:2024qcv,Zhu:2024rej,Redondo-Yuste:2023ipg,Bourg:2024jme,Lagos:2024ekd,Perrone:2023jzq,Kehagias:2024sgh,Bucciotti:2025rxa,bourg2025quadraticquasinormalmodesnull,BenAchour:2024skv,Kehagias:2025ntm, Kehagias:2025xzm,Kehagias:2025tqi,Perrone:2025zhy,Ianniccari:2025avm,Ma:2025rnv}.
These studies have revealed phenomena beyond the scope of linear theory addressing the properties of second-order QNMs and new families of power-law tails. 

While considerable attention has been given in reproducing analytically the numerical results for the nonlinearities popping out from the coalescence of BH binaries in quasi-circular orbits, basically no consideration has been devoted to the nonlinearities generated by ultra-relativistic head-on collisions of equal-mass, non-spinning BHs with different boosts. 
This  might be understandable as the high-energy head-on collision of two BHs is thought to be a highly nonlinear event, possibly not tractable by a simple second-order approach.

In the  standard decomposition in Regge–Wheeler and Zerilli perturbation theory, the ringdown GW strain of a Schwarzschild BH can be schematically be written as 
\begin{equation}
    rh(t,\theta,\phi)= \sum_{\ell m n}A_{\ell m n}  e^{i \omega_{\ell m n }(x-t)}{}_{-2}Y_{\ell m}(\theta,\phi),
    \label{hh}
\end{equation}
where  $A_{\ell mn}$ is the amplitude of the mode $(n,\ell,m)$, ${}_{-2}Y_{\ell m}(\theta)$ are the spin-weighted spherical harmonics and $x=r+2M\ln(r/2M-1)$ ($M$ being the mass of the resulting BH), which describe the angular pattern of the wave and are the natural basis used for the extraction of the mode amplitudes \cite{Cheung:2022rbm}.
 
In  the head-on collision of two BHs the  linear 200 mode dominates the ringdown of the nonspinning Schwarzschild remnant \cite{Sperhake:2008ga} and the resulting nonlinear mode $200\times200$ has an amplitude which has been calculated by numerical simulations to be  \cite{Cheung:2022rbm}

\be
A^{\Psi_4}_{200\times 200}=(1.700\pm 0.115)(A^{\Psi_4}_{200})^2,
\ee
quite independently from the initial conditions, above all the boost factor. 
Here $\Psi_4$ is the Weyl scalar, which is related to the GW strain $h$ asymptotically at large distances by the relation
\cite{Cheung:2022rbm}

\be
r\Psi_4=-2 r \ddot{h}.
\ee
Correspondingly, 

\be
\label{f}
A^{h}_{200\times 200}=\frac{\omega_{200}^2}{2}A^{\Psi_4}_{200\times 200}=(0.118\pm 0.008)(A^{h}_{200})^2,
\ee
where we have used the value $G_NM\omega_{200}\simeq 0.3737$ for the QNM frequency \cite{Berti:2005ys}.
To the best of our knowledge, this numerical result has never been derived analytically. 
We report here the same ratio for the coalescence of BH binaries in quasi-circular orbits resulting into a spinless Schwarzschild BH \cite{Cheung:2022rbm} (see also Refs. \cite{Mitman:2022qdl,Redondo-Yuste:2023seq})

\be
\label{f1}
A^{h}_{220\times 220}=(0.1637\pm 0.0018)(A^{h}_{220})^2,
\ee
which, as we shall see,  will be useful in the following.

The goal of this Letter is  simple: to derive the nonlinearity (\ref{f}) analytically. 
We achieve this by some logical and rather simple steps.  
The first one involves the Close Limit Approximation (CLA), which  allows to think of  a seemingly untractable highly nonlinear problem as a more standard problem of  second-order BH spectroscopy. 
The second is the independence of the nonlinearity from the initial conditions and spin \cite{Redondo-Yuste:2023seq}.
Let us discuss them in turn.
\begin{figure}
    \centering
    \includegraphics[width=0.4\textwidth]{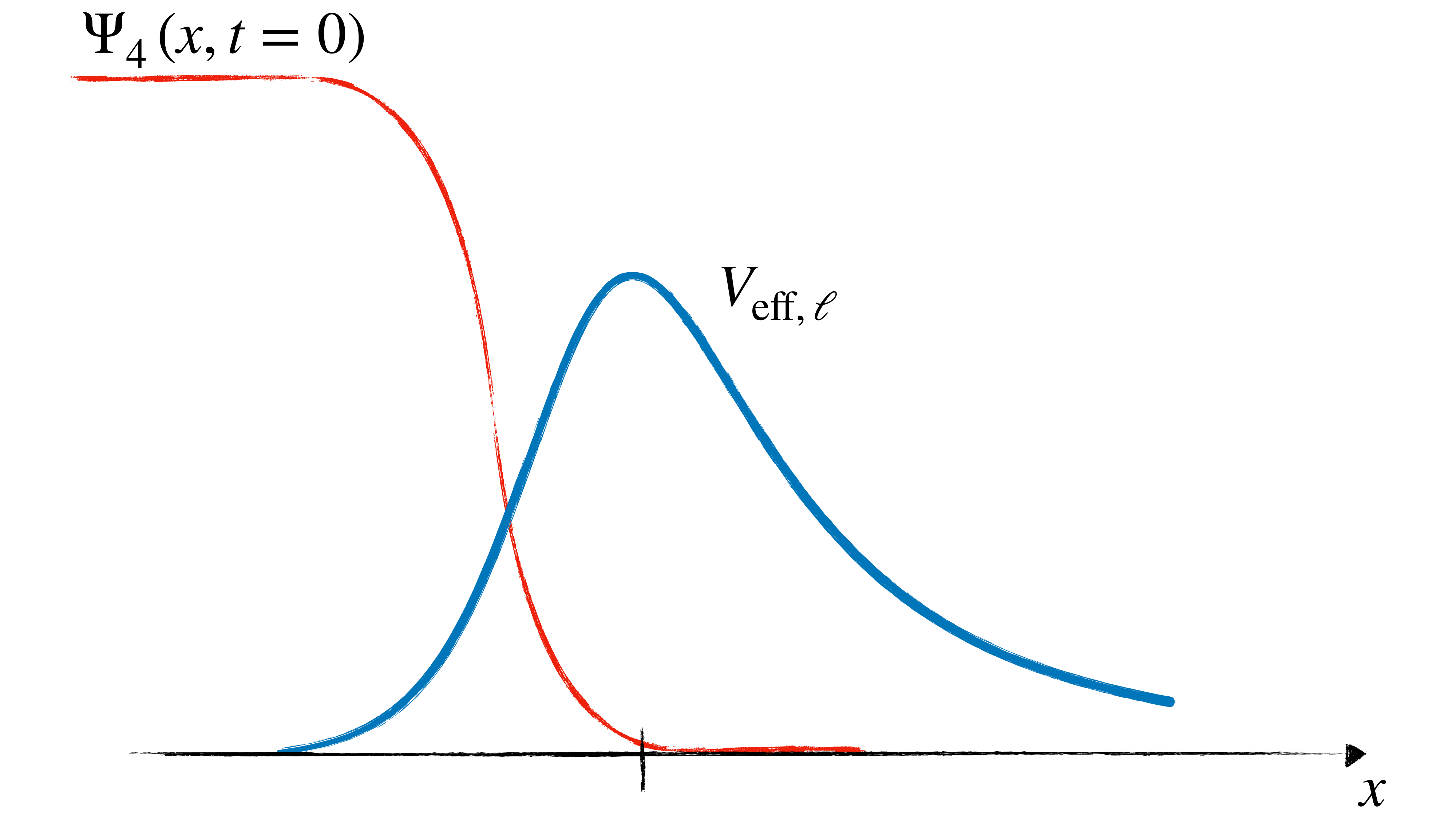}
    \caption{Schematic representation of the initial data and the effective potential of a BH. We used the tortoise coordinate $x=r+2M\ln(r/2M-1)$.}
    \label{fig:init_cond}
\end{figure}
\vskip 0.5cm
\vspace{-5pt}\noindent\textbf{The Close limit approximation --}
The CLA was originally introduced in Ref.~~\cite{Price:1994pm} (see also Ref.~~\cite{Pullin:1997gz} for further analytical developments).
The basic idea is to choose the initial configuration of the two BHs so that they are extremely close, enough to be surrounded by a common apparent horizon. 
From a technical perspective, this means they effectively constitute a single BH. 
Since the apparent horizon already surrounds both BHs, the system can be regarded as a  perturbed version of a single BH spacetime—such as Schwarzschild or Kerr—with appropriate boundary conditions to account for the deformation.

In this CLA, the dynamics of the BH collision is primarily governed by the response of a single BH “ringing down” as it dissipates the distortions caused by the merger. 
Remarkably, the CLA has shown strong agreement with full numerical relativity simulations, even in cases where the expansion parameter is only marginally small \cite{Gleiser:1998rw,Sopuerta:2006wj}.

At first glance, this level of accuracy might seem unexpected, especially considering that the ultra-relativistic head-on collision of two BHs is one of the most nonlinear and complex problems in general relativity. 
The explanation lies in the nature of the initial data: much of it is concentrated within the peak of the Regge–Wheeler or Zerilli potentials $V_{{\rm eff},\ell}$, which govern the linear perturbations of a static BH solution, as depicted schematically in Fig. \ref{fig:init_cond}. 
As a consequence, the bulk of this data falls back into the BH, while only a small portion escapes to infinity.

Therefore, it is natural that second-order perturbation theory performs so well in this context. 
The event horizon effectively “absorbs” the highly nonlinear contributions, confining them to the BH interior, while the weak-field radiative component that reaches the exterior may be accurately described by the second-order theory \cite{Nicasio:1998aj}. 
It is interesting to notice that the same conclusion has been recently obtained and confirmed (empirically) for the case of the QNMs generated by an infalling particle onto a BH \cite{DeAmicis:2024eoy}, where nonlinear effects are small and a description in terms of second-order perturbation theory suffices.

\vskip 0.5cm
\vspace{-5pt}\noindent\textbf{Independence from the initial conditions --} Having argued that perturbation theory is efficient in describing a huge nonlinear phenomenon like the head-on collision of highly boosted BHs, we now draw your attention to the role of the initial conditions. 
Second-order calculations in BH  perturbation theory have shown that nonlinearities are quite independent from the  BH  spin and the initial
data for the perturbations \cite{Perrone:2023jzq,Redondo-Yuste:2023seq}, while numerical simulations have shown independence from the boost factor in the head-on collisions \cite{Cheung:2022rbm}. This can be understood by the following arguments. A second-order perturbation is originated by an equation of the type

\be
{\cal O}_{-2}[\Psi_4^{(2)}]={\cal S}[(\Psi_4^{(1)})^2],
\ee
where ${\cal O}_{-2}$ is the linear Teukolsky operator  and  ${\cal S}$ is the second-order source. 

The solution is the sum of two pieces. 
The first solves the homogeneous equation, however it will not be important for the discussion since the nonlinearities are operatively extracted for second-order perturbations having a frequency which is the double of the one of the first-order perturbation. 
In our case, the frequency extracted is $\omega_{200\times 200}=2\omega_{200}$, which is different from the linear frequency $\omega_{400}$. 

The second piece is the solution of the inhomogeneous equation and it depends on  the second-order source. 
The latter is well peaked, typically around the light ring, where the linear potential has a maximum. 
A WKB approximation then can be applied to show that \cite{Perrone:2023jzq}

\begin{eqnarray}
\Psi_4^{(2)}&=& \left|\frac{\Psi_4^{(2)}(x\rightarrow\infty,2\omega_{200})}{(\Psi_4^{(1)})^2(x\rightarrow\infty,\omega_{200})}\right|\nonumber\\
&\simeq&\left|\frac{1}{2\omega_{200}}\frac{c_2}{c_1^2}
\frac{\Psi_{4+}^{(1)}(x\rightarrow\infty,2\omega_{200})}{(\Psi_{4-}^{(1)})^2(x\rightarrow\infty,\omega_{200})}
\right|,
\end{eqnarray}
where  $\Psi_{4\pm}^{(1)} $ are  the solutions of the homogeneous equation with QNM 
 boundary conditions at $x\rightarrow \pm \infty$. 
 Above all,

 \be
c_2=\frac{c_1^2}{W(2\omega_{200})}\int_{-\infty}^{\infty}{\rm d}x'\,\Psi_{4-}^{(1)}(x',\omega_{200})h_{200\times 200}(x'),
 \ee
where $W$ is the Wronskian of the linear solutions and $h_{200\times 200}$ is defined through the relation

\be
{\cal S}_{200\times 200}(x,\omega)=\frac{i c_1^2}{\omega-2\omega_{200}}h_{200\times 200}(x).
\ee
The independence from the initial conditions is manifest from the cancellation of the coefficient $c_1$ which parametrizes the amplitude of the dominant mode, 

\be
\Psi_4^{(1)}(x\rightarrow\infty,t)\simeq c_1 \Psi_{4-}^{(1)}(x\rightarrow\infty,\omega_{200})e^{-i\omega_{200}t}
\ee
asymptotically far from the BH remnant, where the nonlinearities are measured.
\vskip 0.5cm
\vspace{-5pt}\noindent\textbf{Calculating the nonlinearity for the head-on collision --}
Having offered an analytical explanation of the independence from  the initial conditions of the nonlinearities coming out from the merger of two (making BH remnants from circular orbits or head-on collisions basically equivalent), we are now ready to explain the nonlinear result (\ref{f}). 
The logic is the following. 
Consider a generic second-order source ${\cal S}[(\Psi_4^{(1)})^2]$ {\it before} having performed the multipole decomposition. 
Clearly, this source is universal and only the multipole selection makes it different multipole by multipole. 
Since the source is quadratic
in the fields, the selection of a given $(\ell,m)$ mode brings the Gaunt integral

\begin{eqnarray}
&&\hspace{-0.6cm}\int\hspace{-0.1cm}{\rm d}\Omega{}_{-2}Y_{\ell_1 m_1}{}_{-2}Y_{\ell_2 m_2}{}_{-2}Y^*_{\ell m}\hspace{-0.2cm}=\hspace{-0.1cm}\sqrt{\frac{(2\ell_1+1)(2\ell_2+1)(2\ell+1)}{4\pi}}\nonumber\\
&\cdot&\left(\begin{array}{ccc}\ell_1 &\ell_2&\ell\\
-2&-2&-2\end{array}\right)\left(\begin{array}{ccc}\ell_1 &\ell_2&\ell\\
m_1&m_2&-m_1-m_2\end{array}\right),\nonumber\\
&&
\end{eqnarray}
where the fields in the source have been decomposed in the multipoles $(\ell_1,m_1)$ and $(\ell_2,m_2)$. 

Let us consider the quadratic QNMs with multipoles $220\times 220$ whose nonlinearity is given in Eq. (\ref{f1}) and the $200\times 200$ whose nonlinearity is given in Eq. (\ref{f}). The differences between the two corresponding sources do not arise  from the respective frequencies because $\omega_{200}=\omega_{220}$  for non-spinning  Schwarzschild BHs due to the rotational symmetry they enjoy. 
The QNM spectrum resulting from a non-spinning BH does not depend on the azimuthal number $m$ related to the azimuthal angle $\phi$. 
Since the second-order amplitude $A^h_{\ell_1 m_10\times \ell_2 m_20}$ is an SO(3) singlet, it depends on $m\!=\!-\!m_1\!-\!m_2$ only through the $3j$-Wigner symbol 
$
\begin{pmatrix}
\ell_1 & \ell_2 & \ell \\
m_1 & m_2 & m
\end{pmatrix}.
$
In other words, the $m$
dependence of the quadratic QNMs is fully captured by the $3j$-Wigner symbol and the relative difference between the $200\times 200$ and $220\times 220$ modes is fully captured by the ratio


\be
\frac{\left(\begin{array}{ccc}2 &2&4\\
0&0&0\end{array}\right)}{\left(\begin{array}{ccc}2 &2&4\\
2&2&-4\end{array}\right)}=\frac{\sqrt{2/35}}{1/3}\simeq 0.717.
\ee
With this reasoning, we arrive at the simple expression  

\be
\frac{A^{h}_{200\times 200}}{(A^{h}_{200})^2}\simeq 
\frac{A^{h}_{220\times 220}}{(A^{h}_{220})^2}\cdot \frac{\sqrt{2/35}}{1/3}\simeq 0.1637\cdot 0.717\simeq 0.117,
\ee
which reproduces the value (\ref{f}) found numerically in Ref. \cite{Cheung:2022rbm} remarkably well. 
In the familiar language of high energy theory, our result makes use of a ``bootstrapping" procedure which allows to calculate a given quadratic QNM for spinless BHs given another one. 
As simple it might seem, this result looks  non trivial to us, being the consequence of non obvious logical steps: 
 the realization  that even large nonlinearities may be  efficiently described by a second-order theory and the fact that the way the final BH is formed does depend only mildly from its parents.


\vspace{5pt}\noindent\textbf{Conclusions --} In this note, we have derived the nonlinearity associated to the quadratic QNM for the fundamental mode generated by the ultra-relativistic head-on collision of two BHs and shown it reproduces the numerical result. 
Our derivation relied on the fact that a second-order calculation, based on the CLA argument, suffices and that the dependence on the initial conditions is mild due to the fact that the second-order source is highly peaked around the light ring. 
We made use of the  fact that the energy spectrum of the QNMs is degenerate for a Schwarzschild BH when changing the azimuthal  number $m$. 
Our reasoning therefore will fail in the case of highly spinning BHs for which the QNM spectrum has a stronger dependence on $m$. 
However, the  reasoning about the small impact of large nonlinearities due to a common horizon envelope smearing them out  goes beyond the example we have discussed. 
This argument represents  a generic ``bootstrapping" procedure that may help fixing   other interaction couplings given one.

\vskip 0.1cm
\centerline{\it Acknowledgments}
  \vskip 0.1cm
  \noindent
   We thank G. Carullo for useful comments on the draft. The work of D.P. is supported by the Swiss National Science Foundation under grants no. 200021-205016 and PP00P2-206149.
A.R.  acknowledges support from the  Swiss National Science Foundation (project number CRSII5\_213497) and from  the Boninchi Foundation through  the project ``PBHs in the Era of GW Astronomy''.

\bibliography{main}

\end{document}